# Conformation-Induced Stiffening Effect of Crosslinked Polymer Thin Films

Zhengyang Zhang [a], Pei Bai [a], Yuhan Xiao [a], Yunlong Guo [a,*], Yanming Wang [a,*]

[a] University of Michigan–Shanghai Jiao Tong University Joint Institute, Shanghai Jiao Tong University, 800 Dong Chuan Road, Minhang District, Shanghai 200240, China

Email: yunlong.guo@sjtu.edu.cn; yanming.wang@sjtu.edu.cn

## Abstract

Nanoscale polymeric thin films are widely used in diverse applications such as energy devices, flexible electronics and biosensors, where a satisfactory mechanical performance is of vital importance to realize their full functionality. It has been evidenced that the elastic properties of polymer films are often strongly affected by their thickness; however, the underlying mechanism of this phenomenon, especially a thorough understanding at the microscopic level, has yet to be achieved. Here we established a coarse-grained molecular dynamics (CGMD) based computational framework, combining with experimental verifications, aiming to reveal the conformational origin of the stiffening behavior of crosslinked polymeric thin films. By imposing systematic controls over the polymer network structures, we found that the bi-axial modulus changes are essentially consequent of the alteration of polymer conformations. A unified theory was then proposed, to quantitatively clarify the correlation between the elastic properties of the system and the distributional variations of the chain end-to-end distances, with predicting a significant hardening effect on top of the conventional entropic elasticity with largely uncoiled chains. Adopting processing protocols inspired by the modeling, our experiments showed that PDMS films at approximately the same thickness may exhibit a two order of magnitude difference in their moduli. The good agreement between experiments and simulations illustrated our findings as an effective guideline for tailoring the elastic properties of polymer films at nanoscale.

**Main**

**Introduction**

Polymeric thin films are widely used in a broad range of applications including energy devices[1], transistors[2] and nanocomposites[3,4]. For instance, polymer semiconductors have certain advantages in wearable electronics, thanks to their high flexibility, stretchability and crack resistance[5,6]. Or in nanofabrication, polymer films often act as nanoscale pattern generators by maintaining their topological pattern in prestressing and constrained shrinking[7]. In all these scenarios, the mechanical behaviors of polymer films are considered of vital importance to meet their functional requirements, especially when the thickness of the films shrinks to nanoscale. It has been reported that under nanoconfinement, the mechanical properties, such as the elastic moduli of the polymeric films, drastically change with the decreasing of film thickness[8–12]. To carefully examine these effects, tremendous experimental efforts have been made to measure the mechanical responses at nanoscale by techniques such as nanoindentation[13–15], surface buckling[16], uniaxial tension[17], capillary wrinkling[18] and micro vibration[19], for both freestanding films and those in contact with substrates. However, to date, how the mechanical properties of a polymeric film are related to its intrinsic microstructure is still largely veiled in secrecy. For example, both positive[17,20–22] and negative[23,24] relationships between the film thickness and mechanical behavior have been seen in different polymer systems. Though several factors, including interfacial mobility[25,26], substrate texture[17,19,27,28], and processing procedures[29], have been proposed to explain these correlations, the exact linkage between the change of material mechanical property and the alteration of its intrinsic microscopic properties has not been established. This is mainly hindered by the difficulty in direct observation of polymer chains, originated from the complex and random nature of their conformation. As an alternative approach, in recent years, with the rapid development of software algorithms and hardware facilities, computational techniques such as Monte Carlo simulations[30], classical molecular dynamics[31], and coarse-grained molecular dynamics[32–35], have started to play a more important role in predicting mechanical properties of polymer films, as an effective means to describe the microstructures of polymer network.

In this paper, we aim to trace the microscopic origin hidden behind the thickness dependent elastic properties of crosslinked polymeric system, in aspects of the intrinsic conformational characters. Balancing between the computing efficiency and model accuracy, we developed a coarse-grained molecular dynamics (CGMD) framework containing a set of macro and micro descriptors to comprehensively investigate the elastic behaviors of the polymer system. Based on the results produced by the aforementioned framework, a universal scaling law is established, where the bi-axial modulus is expressed as a function of the distribution of end-to-end distances, with an additional effort to extend the contributions of traditional entropic elasticity by a conformation-governed hardening effect. Finally, adopting model-guided spin coating protocols, we fabricated crosslinked PDMS films with different thicknesses, followed by elastic modulus measurements using an in-house micro vibration system. The results show that the PDMS films made from different fabrication pathways, though at approximately the same thickness, may exhibit up to a two order of magnitude difference in moduli. This again suggests that the drastic change in elastic responses of polymeric systems is expected to originate from the conformational alterations, further than a nanoconfinement effect[15,35] or a surface tension induction[25] elaborated in literatures.

**Results**

**Elastic moduli tuned by conformational control**

Crosslinked polymeric systems were constructed based on the Kremer-Grest model[33], on top of which various constraints were imposed to describe a microstructure design space of the materials. The molecular density $\rho$=0.85 g/cm$^3$ and the crosslinking density of 6.67% (see Table S1 for conversion rule and Table S2 for crosslinking methods in molecular simulations) were kept the same across all the polymeric systems, including both the thin film configurations and the reference bulk one. Besides, all the chains were fully connected (i.e. each tri-functional crosslinking agent formed three bonds) to eliminate the influence caused by unsaturated crosslinking (as it is obvious that with the increase of crosslink density, the stiffness of the polymeric system will increase[25], see Figure S1).

As shown in Figure 1, our design space is composed of several key variables, to systematically examine the structure of crosslinked polymers. These variables, both macroscopic and microscopic, are briefly explained as follows. Firstly, the thickness of the polymer, denoted as $\lambda$, is considered for both freestanding films and those placed on substrates (Figure 1a). Then, the length of individual chains (Figure 1b), controlled by a harmonic potential added to both ends of a chain (after crosslinking, it becomes a segment of the polymer network), is included and represented by a pre-defined uncoiling factor $\alpha$, ranging from 0 (when all monomers collapse to one point) to 1 (when the chain is perfectly straight). Next, the linking pattern of the polymer network is chosen as a variable, which is illustrated using the graph theory with the nodes (CG beads) connected by undirected edges (CG bonds) (Figure 1c). With this degree of freedom, distinguishment can be made between uncontrolled dynamical crosslinking (random connections of crosslinkers to mimic real synthesis operations) and engineered crosslinking patterns (ordered connections of crosslinkers theoretically synthesizable by extreme precision techniques). It should be noted that in our framework, many of the polymer systems are set to have the same number of monomers per individual chain $N$ (e.g., $N = 35$). But for certain samples, the effects brought by non-uniform $N$ are examined. As shown in Figure 1d, this is achieved by assigning $N$ to follow a Gaussian distribution, where its variance $\sigma$ varies with a fixed mean $\mu$ (e.g., $\mu = N = 35$). In addition, the dynamic crosslinking process of the polymers can be investigated by the CGMD model, with considering the variables of the film thickness $\lambda$ and the substrate rotating speed $\omega$ (Figure 1e). More detailed modelling information can be found in the Method Section and Supplementary Information.

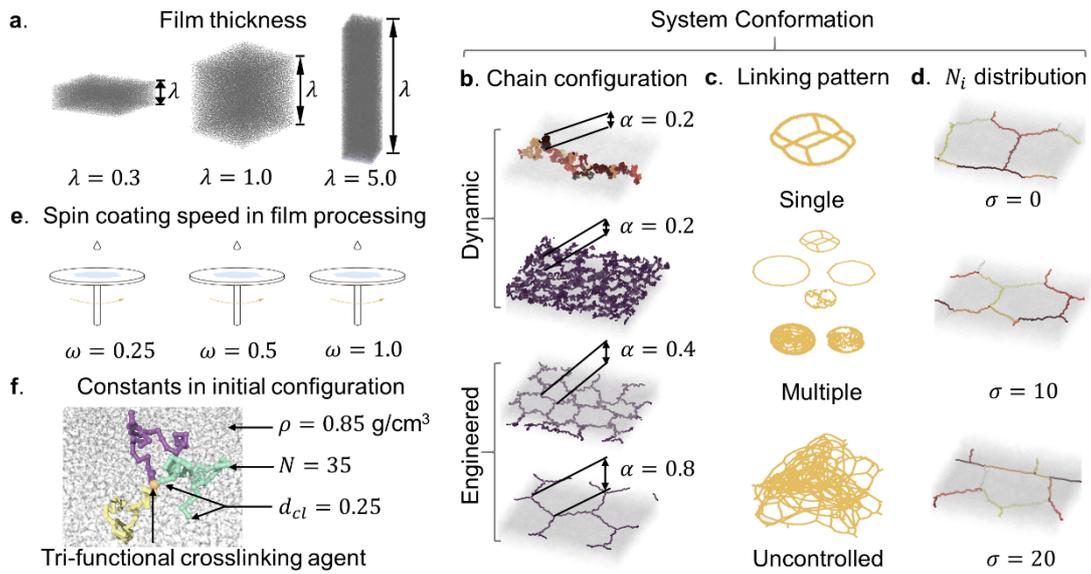

**Figure 1. Schematic of the design space and controlled parameters of polymeric films in the CGMD framework. a**, Thickness of polymers in freestanding or with-substrate status: from less than the gyration of a single chain to bulk by assigning the periodic boundary condition along *z* direction. **b**, Conformations of individual polymer chains: from uncontrolled heavily coiled chains to largely protracted chains. **c**, Crosslinking pattern: dynamic crosslinking with no control, and designed crosslinking patterns with different combinations of individual chain conformations. **d**, Segmental repeating unit number: from a unified number to a Gaussian distribution with different standard deviation. **e**, Spin coating speed: a factor controlling the processing procedure.

In each CGMD simulation, the polymer was first equilibrated adopting a two-step protocol[36], followed by subsequent conformational control, crosslinking, relaxation and biaxial in-plane tensile deformation. For all the polymeric systems, the conformational parameters are obtained before tensile deformation (after crosslinking and relaxation). The bi-axial moduli were estimated using the data in a linear strain-stress regime, to eliminate potential ambiguities caused by nonlinear or rate-dependent behaviors (Figure S2).

In Figure 2a, it can be seen that a pure decrease in thickness $\lambda$ could moderately enhance the stiffness of a polymeric system with a factor of less than 3, which may be related to a pronounced surface tension effect. However, the predicted degree of modulus increasing was far less than a maximum of around 100 times in experiments[19,37], thus, a more detailed investigation awaits to be proposed.

Leveraging the flexibility of the CGMD framework, the conformational features of the polymer system and their correlations to elastic modulus were carefully examined in Figure 2b-e. A quasi-linear relationship of the biaxial moduli with the average uncoiling factor $\alpha_h$ was observed in Figure 2b, where $\alpha_h$ was controlled by adding an external harmonic potential. As it was observed that the above conformational perturbation method became ineffective when $\alpha_h$ was increased from 0.2 (the unperturbed case) to 0.4, a direct control approach was implemented to freely change the uncoiling factor ($\alpha_d \in (0, 1)$) by scaling the simulation cell of an orderly crosslinked polymer system. Interestingly, here the biaxial moduli markedly increased (by over 100 times) at a large uncoiling factor ($\alpha_d = 0.8$) (Figure 2c). Then, a natural further step is to investigate how this stiffening effect depends on the degree of uncoiling uniformity. Along with this direction, polymer chains with different $\alpha_d$s were stacked to form one polymer system at a given mixing ratio. As shown in Figure 2d, a two order of magnitude enhancement of elastic moduli could still be achieved, with only 30% of the chains at $\alpha_d$ of 0.8, while the conformation of the rest 70% of the chains are similar with that in bulk polymer (i.e. $\alpha_d \approx 0.2$). The non-uniformity of the distribution of $N$ (the number of beads per chain segment) is also considered by adjusting its variance $\sigma$. In this case, the bi-axial moduli show a positive relationship with $\sigma$ in the range from 5 to 30 (Figure 2e). This indicates that chains with different conformations (e.g., quantified by $\alpha$) should make unequal contributions to the mechanical properties of the polymer system. Then, one interesting question could be raised: whether this conformation related stiffening behavior can be elucidated by one single descriptor, by aggregating the information from all the above factors.

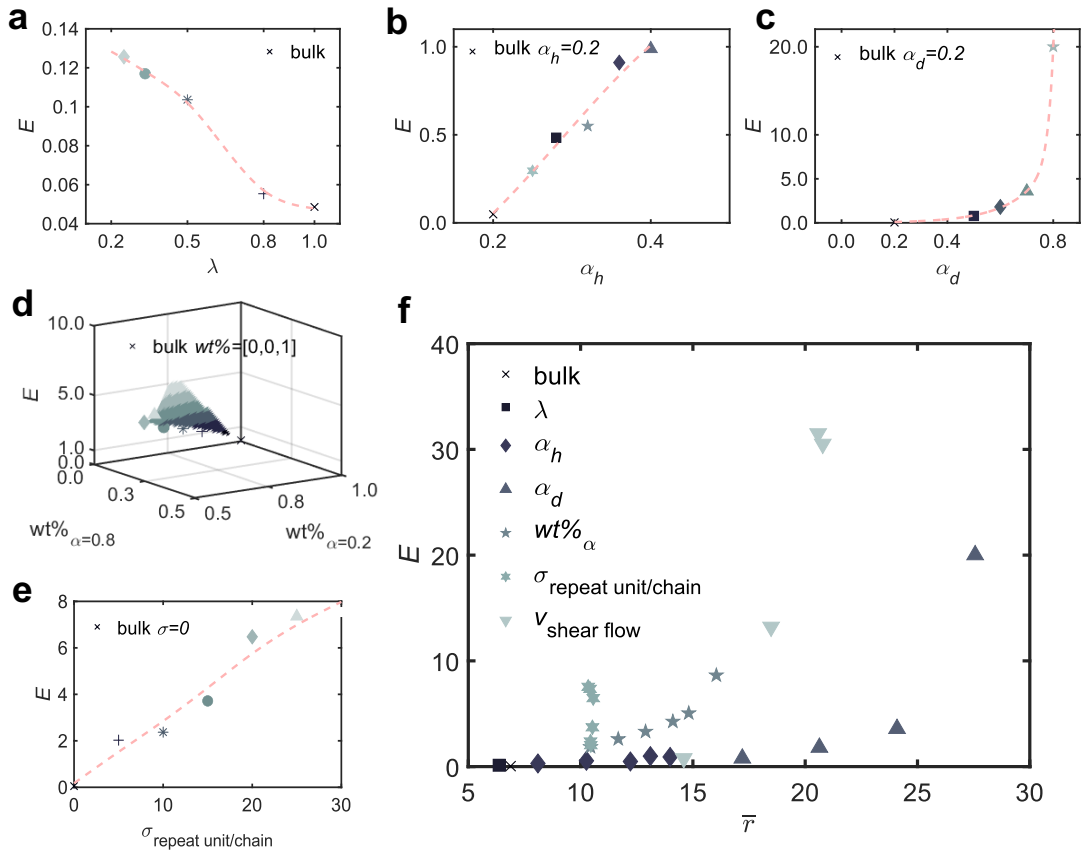

**Figure 2. Controlled paraments with bi-axial Moduli *E*. a**, Thickness $\lambda$ vs *E*. **b**, Intensity $P_I$ and distance $P_r$ of uncoil harmonic potential vs *E*. **c**, Uncoiling factor $\alpha$ vs *E*. **d**, Weight ratios of different pre-protracted chains with $\alpha$ = 0.2, 0.4 and 0.8. **e**, Standard deviations of the distribution of bead per chain $\sigma$ vs *E*. **f**, Mean conformational descriptor $\bar{r}$ as a sign of moduli *E* in each system: ✘ bulk polymer; ■ thickness $\lambda$; ♦ uncoiling factors $\alpha_h$ by harmonic potential; ▲ uncoiling factor $\alpha_d$ by pre-defined chain monomers and crosslinkers; ★ weight ratio of chains with $\alpha_d$ combinations; ✱ randomness of monomer number per chain σ; ▼ spin coating speed *v*;

## Distribution of end-to-end distance as an effective descriptor

Theoretically, the mean end-to-end separation distance ($\bar{r}$) has long been considered as an important descriptor of polymer conformations, whose change could largely affect the mechanical properties of the material[38]. For example, several classic viscoelastic mechanism models, such as the Rouse model[39] and the Ngai coupling model[40], have related $\bar{r}$ to the relaxation time and the compliance of the polymer. Thus, as shown in Figure 2f, $\bar{r}$ was estimated for each of the polymeric systems that were previously constructed in our design space with different structural parameters and processing conditions, and was then plotted against the corresponding bi-axial modulus $E$ (the relationship between the mean gyration vector $\bar{G}$ and $E$ was also tested and results are shown in Figure S3). While a positive correlation can be generally seen between $\bar{r}$ and $E$, the highly scattered data points suggest that this crosslinked polymeric system could neither be treated as ideal chains nor predicted by ideal rubbery elasticity theories. In other words, $\bar{r}$ by itself may be inadequate to serve as a "universal" descriptor, especially for the case where a drastic change in $E$ was accompanied with nearly no change in $\bar{r}$ (the points marked by ✶).

We speculate that the failure of $\bar{r}$ is due to the fact that these crosslinked polymer thin films, engineered by either direct structure designs or processing parameter controls, may result in a distribution of $r$ significantly deviated from a Gaussian-like distribution, the one typically assumed for bulk systems. Thus, for different polymeric systems, it would be of interest to visualize the distributions of $r$. As expected, various distinguishable features were observed on the histograms of $r_i$ for individual chains, including drifting (Figure 3a and b), splitting (Figure 3d and e) and flattening (Figure 3f) of the peaks.

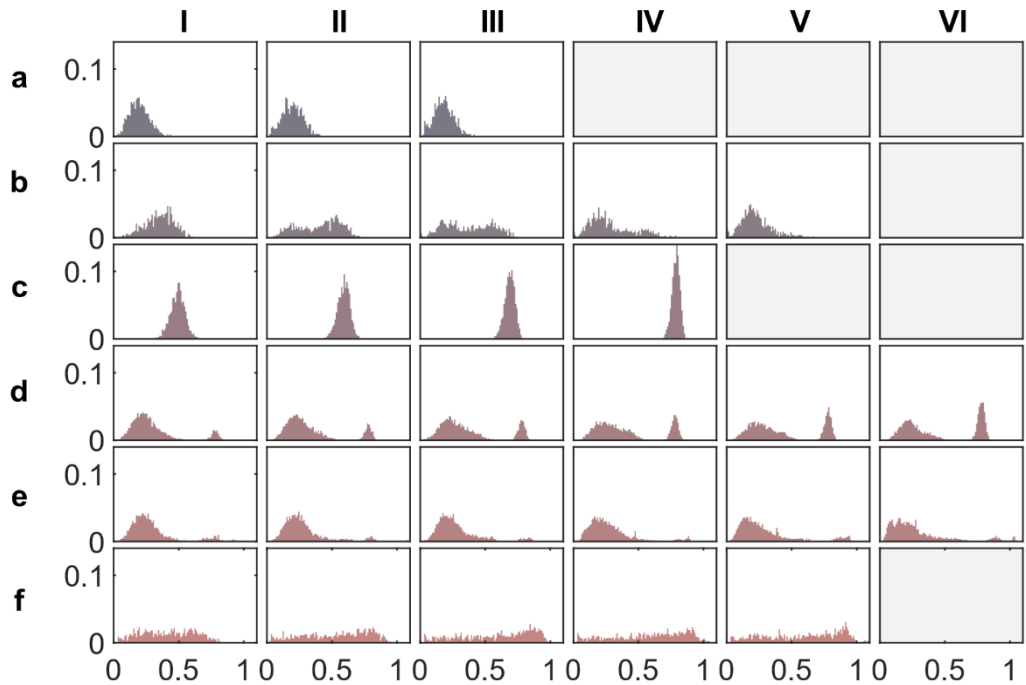

**Figure 3. The distributions of *r* of different polymeric systems before and after deformation. a**, Bulk polymer, **a-I**, Bulk cross-linked polymer **a-II**, Thin cross-linked polymer **a-III**. **b**, Crosslinked polymer under uncoil harmonic potential: **b-I**, Peak at around 17; **b-II**, Peak at around 21; **b-III**, Peak at around 24; **b-IV**, Peak at around 28. (Peak shifts to the right) **c**, Designed cross-linked polymer structure with a single-peak skewed distribution of *r*. **d**, Designed cross-linked polymer structure with a two-peak distribution of *r*: decrease of left peak height and increase of right peak height from **d-I** to **d-VI**. **e**, Designed crosslinked polymer with random segmental bead number per chain shows a three-peak distribution of *r*. **f**, Distribution of *r* of the polymer prepared by shear-crosslinking. The distribution is flattened by shear in processing. The peak moves from left (**f-I**) to right (**f-V**) with the increase of shear speed.

**Extended scaling theory for in-plane stiffness of crosslinked polymer**

For a typical bulk elastomer system, the elastic modulus $E$ normally scales with $-3k_BT/Nb^2$, where $N$ is the number of Kuhn segment and $b$ is the Kuhn length. Obviously, the above model can neither explain the positive relationship between $E$ and $\bar{r}$ nor include the influence of altered $r$ distributions. Here we hypothesize that in crosslinked polymeric thin film containing a noticeable amount of uncoiled chains, additional stiffening effect may need to be considered, with its magnitude positively correlated with both the number of uncoiled chains and the uncoiling degree of these chains. In the CGMD simulations, the above information can be obtained by calculating the end-to-end distance for individual chains (denoted as $r_i$) and counting the number of beads of chain *i* (denoted as $N_i$). Thus, a two-term universal scaling law can be proposed for explaining the simulation data: the first term is based on the classical model that sets the entropic contribution inversely proportional to $r_i^2$; while the second term accounts for a 'hardening effect', which is expected to have a positive relationship with $r_i$. In addition, observed from Figure 2, this term should be negligible for bulk systems (i.e. when $r_i$ is relatively small) and quickly rise up at some specific $r_i$ value, exhibiting a behavior coincident with that of an activation function in artificial neural network (ANN) models (e.g., a customized hyperbolic tangent function, see Figure S5). Thus, the relationship of the biaxial moduli and $r_i$ can be expressed as:

$$E \cong \frac{1}{n}\left[\varphi \sum_{i=1}^{n} \frac{N_i^2}{(\overline{N}r_i)^2} + \theta \sum_{i=1}^{n}\left[1 + \tanh\left(p\frac{r_i}{N_i} - q\right)\right]\right] \quad \text{(Equation 1)}$$

where $n$ is the total number of chains in the system, $N_i$ is the number of beads (repeat units) per chain segment, $\overline{N}$ is the average number of beads per chain segment (in this case, $\overline{N} = 35$), $\varphi$ and $\theta$ are the two scaling parameters tuning the stiffness of the two terms respectively, and the shape of the tanh function is controlled by the other two fitting parameters $p$ and $q$. When at a large enough $n$, the distribution of $r$ may be approximated by a continuous probability density function $\rho(r)$. Then, the $E$-$r$ relationship may be written as $E \cong \int \left[\varphi \frac{N_r^2}{(\overline{N}r)^2} \rho(r) + \theta \left(1 + \tanh\left(\frac{pr}{N_r} - q\right)\right) \rho(r)\right] dr$.

The above scaling law makes consistent predictions on the CGMD simulated biaxial moduli of all the polymer configurations, while clear divergence was observed when comparing the predictions by a classical model with the CGMD data (Figure 4a). According to the scaling result, the conventional entropic elasticity (the first term) still holds its dominating position on the elasticity for bulk polymeric system, when the uncoiling factor $\alpha = \frac{r}{N}$ ~0.2-0.3. An increase of $\alpha$, especially when $\alpha$ is larger than 0.7, leads to a drastic rising of the hardening effect term (the second term) with the diminishing of the first term, presumably as the major contribution to the over-two-magnitude difference between the bulk and the crosslinked polymeric nano-film (Figure 4b).

All the examined polymer systems were well equilibrated before the biaxial tensile tests. Despite the conformational change from a bulk polymeric system, the bond distances and bond energies barely changed (e.g., < 0.2% in Table S3 and Figure S4). This affirms that the stiffness change of the crosslinked polymer should be mainly attributed to the alteration of chain conformations, rather than the stretching of individual chemical bonds. Certainly, the relative stiffness of the hardening effect (that may be represented by the intersection point of the first term and second term) may depend on the crosslink density, chain length, and etc. For instance, a higher crosslink density could result in an earlier intersection. In addition, the mechanical performance of crosslinked polymers should be affected by their connecting modes. The modulus of a polymer system with all the crosslinkers connected in parallel may correspond to an upper limit estimate; the modulus of a perfectly series system may be regarded as a lower bound; and a more general polymer network is expected to have the modulus with its value in between. Further demonstration of series/parallel effect of the polymer chains can be found in the Supplementary Information (Figure S6).

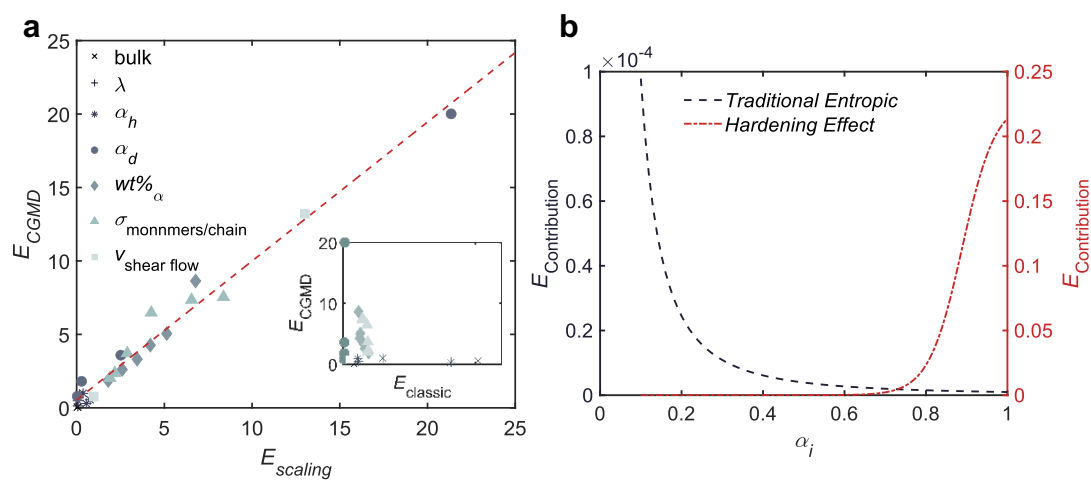

**Figure 4. Scaling results of polymeric systems. a**, fitted biaxial tensile moduli versus raw CGMD data. The fitted line passes origin (0,0). The subfigure presents the results of classic entropic term versus the raw CGMD data, which cannot find clear correlations between each other. **b**, Contribution of first term (classic entropic elasticity) and second term (hardening effect) to the biaxial tensile moduli of polymer chains versus their elongation factors.

**Tailored crosslinking process altering chain conformations**

Evidence has accumulated that the processing procedures, including spin coating and the cure of dynamical crosslinking (normally used to prepare crosslinked thin films like PDMS), could largely change the microstructure and conformation of the polymeric system. To investigate the associated underlying mechanisms, the spin coating and polymer curing processes were simulated by the CGMD framework, specifically considering the influences of parameters such as tangential velocity $v_t$, an effective centrifugal force $f_c$ applied on the chains in the CGMD simulation, and the film thickness $\lambda$ on the polymer microstructures (Figure 5a, b and c).

When the polymer is in contact with a substrate, the simulation results showed that the applied tangential velocity $v_t$ (or the spinning speed $v$ after conversion) positively correlates with $r$. With the increase of the spinning speed, the distribution of $r$ becomes more flattened in contrast to a typical Gaussian distribution (Figure 3f). According to Equation 1, a larger $r$ would lead to a more pronounced hardening effect, finally causing the increase of the bi-axial modulus (Figure 5h). The simulations also reveal the role that the film thickness $\lambda$ plays in affecting the polymer conformations. As shown in Figure 5c, for a thick polymer film ($\lambda=5.0$), only the chains close to the bottom substrate are stretched to a very large extent. Along the $z$ axis, the chains in the middle section of the simulation cell are moderately affected, while the conformation of the chains in the top region almost remains unchanged (in comparison with that of a bulk system) in a statistical perspective. For the polymer film with a $\lambda$ of 1.0, the conformations of the chains are all affected, with their degree of uncoiling descending along the $z$ axis (Figure 5d). However, the chains in an ultra-thin film ($\lambda=0.2$) appear to be 'stuck' to the substrate, with their conformations nearly unchanged (Figure 5e). These phenomena suggest that a mobility gradient of the atom of the polymer chains (analogy to the concept used by Hao, et al.[41]) between the adjacent layers of chains could be generated during the shearing process, which may act as the driving force for the polymer chain conformation changes. As a consequence, the elastic moduli of the system with different thickness presents an increase-before-decrease trend (Figure 5i). From the above discussions, the stiffening of the polymer film is a joint effect of $v_t$ and $\lambda$, which clearly cannot be solely described by each single parameter. (Related discussion can be found in Figure S7 and Table S4).

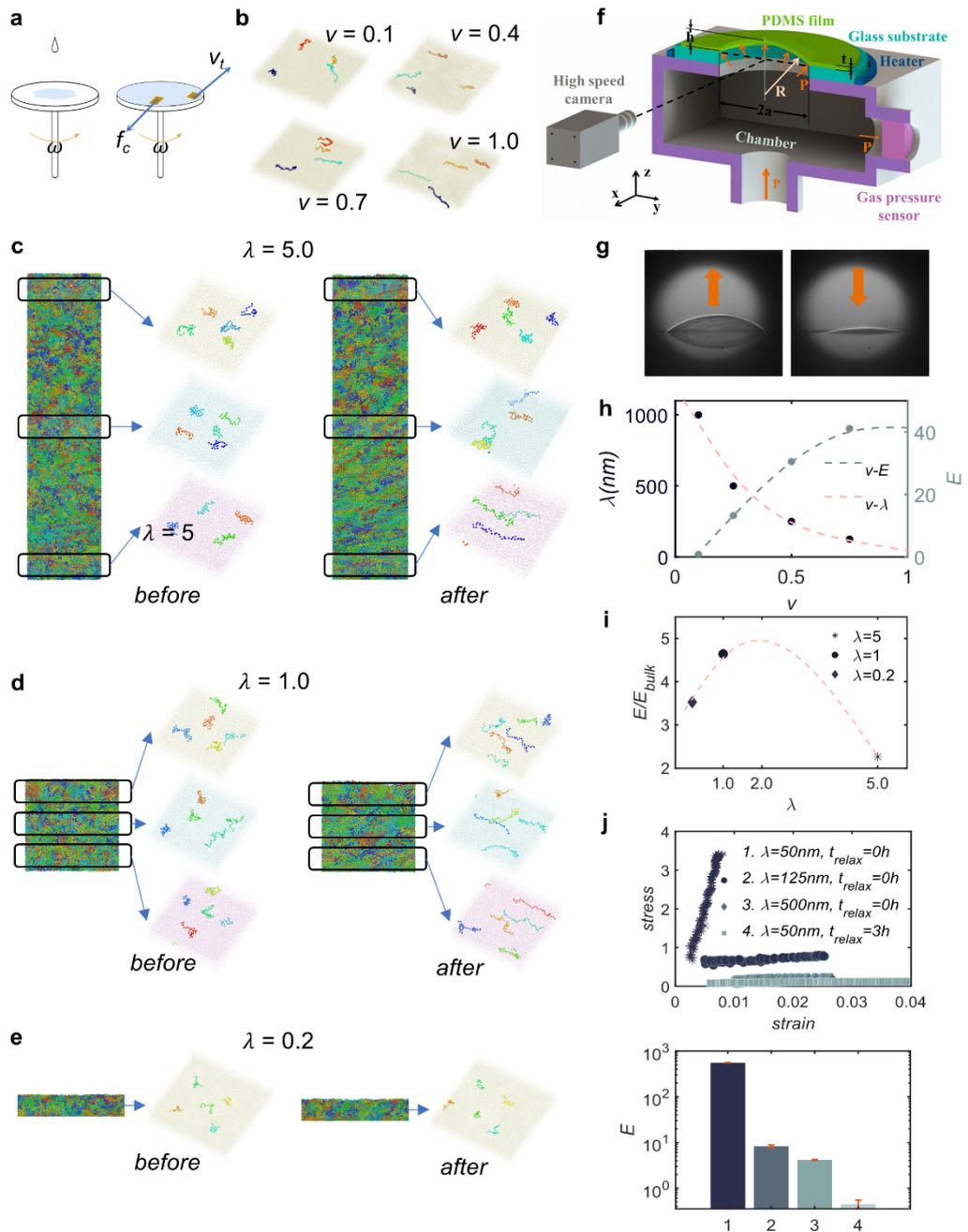

**Figure 5.** Shear-crosslinking of polymer. **a**, Schematic illustration of spin coating with tangential velocity $v_t$. **b**, Conformation change under different spinning velocity derived by spinning speed. **c-e**, Conformation change at different thickness at the same spinning speed represented by centrifugal force $f_c$ (The dynamics of **b-e** could also be found in Figure S7). **f**, schematic illustration of in-house micro-vibrational test device. **g**. photo of the thin polymer films **h**, relationship between spinning velocity $v$, Bi-axial Moduli $E$ and thickness $\lambda$. **i**, relationship between spinning velocity $v$, Moduli $E$ and thickness $\lambda$. **j**, Difference in moduli between spinning coating–relaxing–curing and spinning coating–curing without relaxation.

Under the guidance of the aforementioned simulations and scaling theory, crosslinked

PDMS thin films samples were prepared by a spinning coating-crosslink curing procedure, using tri-functional crosslinking agent to generate connecting topologies comparable with the model. The mechanical properties of the freestanding crosslinked PDMS samples with various film thicknesses were measured on our in-house micro vibration device (Figure 5f, Figure S8). The biaxial moduli of the samples were calculated based on curvature of the photos taken by high-speed camera[19] (Figure 5g, Figure S9, Equation S2 & S3). A dramatical increasing of the biaxial modulus (~135 times) was found when the film thickness decreased from 500 nm to 50 nm, achieving a consistency among our previous experimental work[19], the CGMD predictions and the scaling theory (Figure 5h and Figure 5i). This trend could be explained as follows: since thinner polymer films were fabricated by the application of a higher rotating speed, according to our model, this would lead to more stretched chains during the spin-coating process. Then, followed by an almost immediate curing process, the largely altered chain conformations were preserved by the crosslinking, resulting in the increase of the modulus. To validate this theory, a controlled experiment was then designed, such that a 50 nm thick sample was spin-coated and stood at room temperature for 3 hours, before entering the heat curing stage where nearly all the crosslinkers were activated. In this case, while the thickness of the film did not show noticeable changes, the uncoiled chains naturally tended to recover during the 3 hours relaxation; then presumably the chain conformation in the final crosslinked film would be closer to that of a thick film. If our hypothesis was correct, the biaxial moduli of this controlled sample should be much lower than the value of the 50 nm sample which was immediately cured after spin coating. In Figure 5j, the stress-strain curve of the 50 nm PDMS relaxed sample was experimentally depicted, alongside with those of 50 nm and 500 nm PDMS films. It can be clearly observed that consistently with our expectation, polymer films with both 50nm thickness show disparate elastic performance (with an over two-magnitude of difference) between the case of immediate crosslinking and the case of crosslinking after 3 hours relaxation. The above experiments reemphasized our findings that the thickness dependent elastic properties of crosslinked polymeric thin films should originate from the alteration of chain conformations. In other words, in principle, a decoupling between the film thickness and film stiffness can be achieved, if the chain conformations can be independently controlled.

**Discussion**

This work aimed to link the elastic property of the polymeric films to the distributional information of the chain conformations, by combining a CGMD framework and experimental validations. A two-term scaling law was established to make accurate predictions on crosslinked polymer systems given a distribution of the end-to-end distances, as an important step to unveil the microscopic origin of the stiffening effect of polymer networks. Based on these findings, the investigation on the processing conditions of polymer thin films provided useful guidelines for tailoring their elastic properties for many exciting applications such as wearable electronics and flexible energy devices. To extend the applicability of our framework to more general polymer systems, a further step would be to develop a stable algorithm for including the factor of entanglements , which are expected to be highly correlated with the stiffness, strength, and failure of many long-chain polymers[42]. Also, while the adoption of a FENE potential in the current modeling framework is a comprise to the efficiency and flexibility of the model; for specific polymer systems, a more sophisticated potential, e.g. with the consideration of molecule anisotropy[43], may be needed for accurately describing the intermolecular and polymer-substrate interactions. Hence, the chemical species for both the polymers and the substrates may potentially be added to the parameter space, for designing better polymer films and optimizing their fabrication pathways. In addition, it may be of interest to model the breakage of chemical bonds, to enable predictions of self-healing behaviors[44,45] and/or damages of polymer networks. As a long-term goal, one promising direction is to merge the above components with the work present in this paper, to form an integrated modeling framework applicable to an even wider range of polymer systems containing both long and short chains, with or without crosslinkers. This will allow for the construction of self-consistent and high-quality polymer databases, towards a data-driven approach for engineering polymer structures with tailored mechanical properties.

**Methods**

**Simulations**

Bead-spring model extended from the Kremer and Grest model[33] was used to perform the relaxation, crosslinking and tensile test. The bonds were represented by the Finitely Nonlinear Elastic (FENE) potential while the non-bonded beads interactions were represented by 12-6 Lennard-Jones potential. Crosslinked PDMS films were chosen as a representative crosslinked polymeric system, with the number of beads per chain $N$ kept at 35 for all the systems (according to experiment[46] and previous simulation[33]) except those with σs. The bond formed between crosslinkers and chain ends were considered the same with those between chain beads. All the CGMD simulations were performed via the Large-scale Atomic/Molecular Massively Parallel Simulator (LAMMPS) [47].

For simulating the dynamical crosslinking process, chains and crosslinkers were randomly created with desired molecular density (0.85 g/cm³) and crosslinker density (6.67%). A layer of atoms was created below the polymer to represent the existence of substrates for some of the systems. An initial equilibration of the polymer film was firstly conducted with $10^4$ steps under soft dissipative particle dynamics (DPD) potential with gradually increasing interaction force and then $10^7$ steps under Lennard-Jones (LJ) potential for a well-established structure[32,36]. After that, different conformational controls were implemented as follows. To control $\lambda$, polymer with different thickness were created by direct generation from random positioned chains in orthogonal simulation boxes such as $\lambda=0.2$ ($x=y=5z$), $\lambda=1$ ($x=y=z$) and $\lambda=5$ ($x=y=z/5$) with periodic boundary conditions on x and y direction and non-periodic boundary conditions on z direction. LJ walls were used to confine the polymer chains before they are crosslinked to hold their own thin film structure. Full periodic boundary conditions were applied to simulate bulk polymers with a box of $x=y=z$. To control $\alpha_h$, an external harmonic potential, parametrized by the spring constant $K$ and the equilibrium distance $r_0$, was added to uncoil the chains. In other cases, different velocities were added to the substrate underneath the polymers to simulate the normal speed of the substrate in processing procedures like spin coating. Or external forces were imposed to the polymer chains to simulate the radial forces created by spinning. For each of those controls, enough time was given to make it effective before the chains were crosslinked at the ends by crosslinkers. After the systems were sufficiently crosslinked, the conformational controls including walls, harmonic bonds, velocities and external forces were all removed and a further relaxation of $10^7$ steps under LJ potential was done before deformation test.

For engineered polymer network patterns, crosslinkers and linking bonds were created between pre-location chains to form honeycomb-like structures with same molecular density and crosslinker density to control $\alpha_d$. The combinations of different $\alpha_d$s in one polymer system were also realized by mixing engineered chain networks. For each of the pre-designed polymers, the equilibration protocol was performed before the numerical tensile test.

Groups with different σs from 0 to 30 linearly spaced by 5 were introduced both in dynamical crosslinking polymers and engineered polymers. The biaxial deformation tests were performed at a strain rate of $10^{-4}$, and it was verified that the response of the system was not sensitive to strain rate (Figure S2).

**Experiments**

Silicone Elastomer Kits ( Sylgard® 184 purchased from Dow Corning, USA) was used to fabricate the PDMS films. A Cellulose acetate (CA) with 39.8 wt.% acetyl group (Aladdin, China) served as the sacrificial layer material to peel the film off the glass substrate. The PDMS precursors first were mixed with the crosslinking agent with a ratio of 10:1 wt./wt., then a PDMS mixture/toluene solution (4-10 wt.% solute) was spin-coated onto glass slides with rotating speeds from 2000-4000 rpm for 1 min. After spin-coating the sacrificial layer and the PDMS layer, the film was held at 80 ºC in a vacuum oven for 5 h. For the control experiment, the sample was relaxed at the room temperature for 3 h before the curing step. For the mechanical test, the stress-strain curve was obtained taking the same procedures in our previous work[19], with the equipment information and calculation formula provided by Figure 5g, Figure S9, and Equation S2 & S3.

# Supplementary Information

# Conformation-Induced Stiffening Effect of Crosslinked Polymer Thin Films

Zhengyang Zhang [a], Pei Bai [a], Yuhan Xiao [a], Yunlong Guo [a,*], Yanming Wang [a,*]

[a] University of Michigan–Shanghai Jiao Tong University Joint Institute, Shanghai Jiao Tong University, 800 Dong Chuan Road, Minhang District, Shanghai 200240, China

Email: yunlong.guo@sjtu.edu.cn; yanming.wang@sjtu.edu.cn

**Table S1.** Unit conversion rule of the simulation.

| System | MD | PDMS |
|---|---|---|
| $T$ | $1\epsilon/k_B$ | 300 K |
| Monomer mass | 1 | 74 g/mole |
| $1\sigma$ | $4.5\sigma$ | 8.7 Å |
| $1\tau$ | - | $2.3\times10^{-10}$ s |

**Table S2.** Crosslinking methods in molecular simulations.

|  | PDMS molecule inners | PDMS molecule ends | Crosslinkers |
|---|---|---|---|
| PDMS molecule inners | ✗ | ✗ | ✗ |
| PDMS molecule ends | - | ✗ | ✓ |
| Crosslinkers | - | - | ✗ |

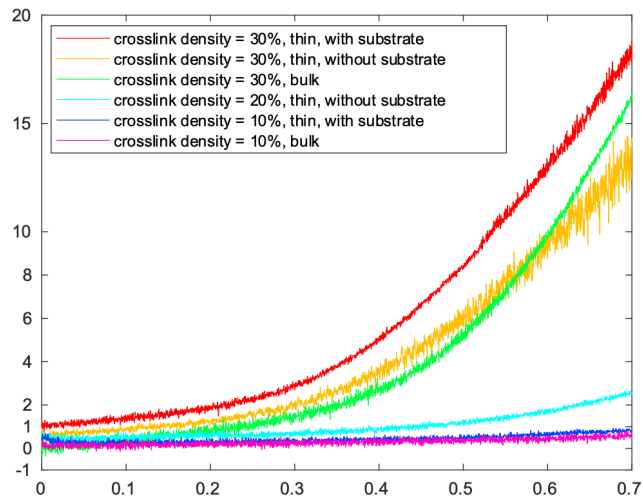

**Figure S1**. The effect of the polymer crosslinker number and the existence of substrate on the stiffness of the polymeric system.

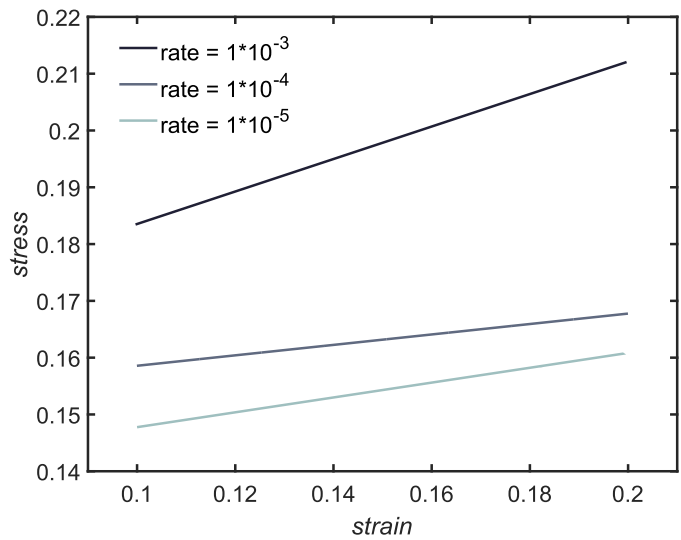

**Figure S2**. The rate dependency of polymer's elastic behavior. Biaxial tensile strain rates were set as $1*10^{-3}$, $1*10^{-4}$ and $1*10^{-5}$. The molecular density, crosslink density and all other parameters are the same as the simulations in this work.

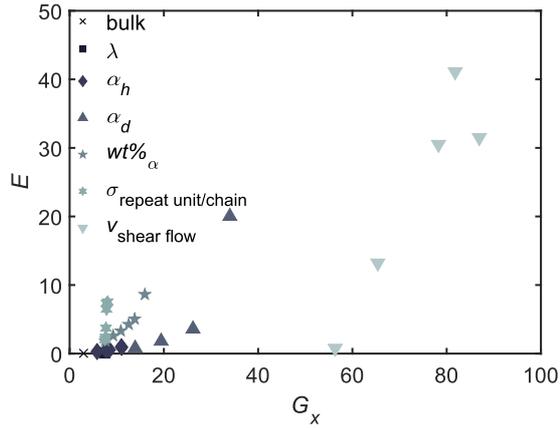

(a)

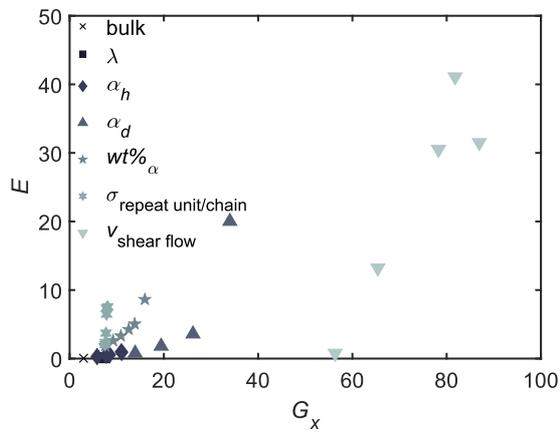

(b)

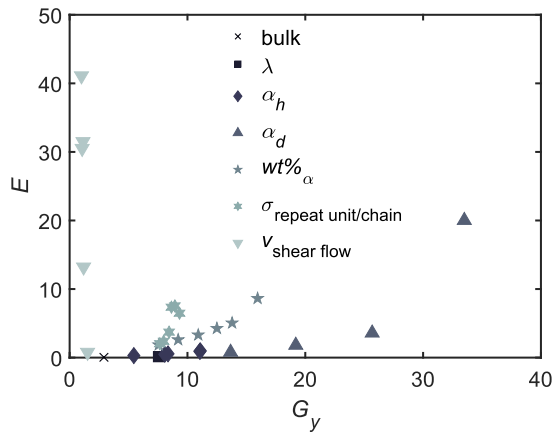

(c)

**Figure S3**. **Controlled paraments with bi-axial Moduli $\bar{G}$ as signs of moduli $E$ in each system**: ✖ bulk polymer; ■ thickness $\lambda$; ♦ uncoiling factors $\alpha_h$ by harmonic potential; ▲ uncoiling factor $\alpha_d$ by pre-defined chain monomers and crosslinkers; ★ weight ratio of chains with $\alpha_d$ combinations; ✱ randomness of monomer number per chain σ; ▼ spin coating speed $v$; (a), (b) and (c) Relation between $\bar{G}_x$, $\bar{G}_y$, $\bar{G}_z$ and $E$.

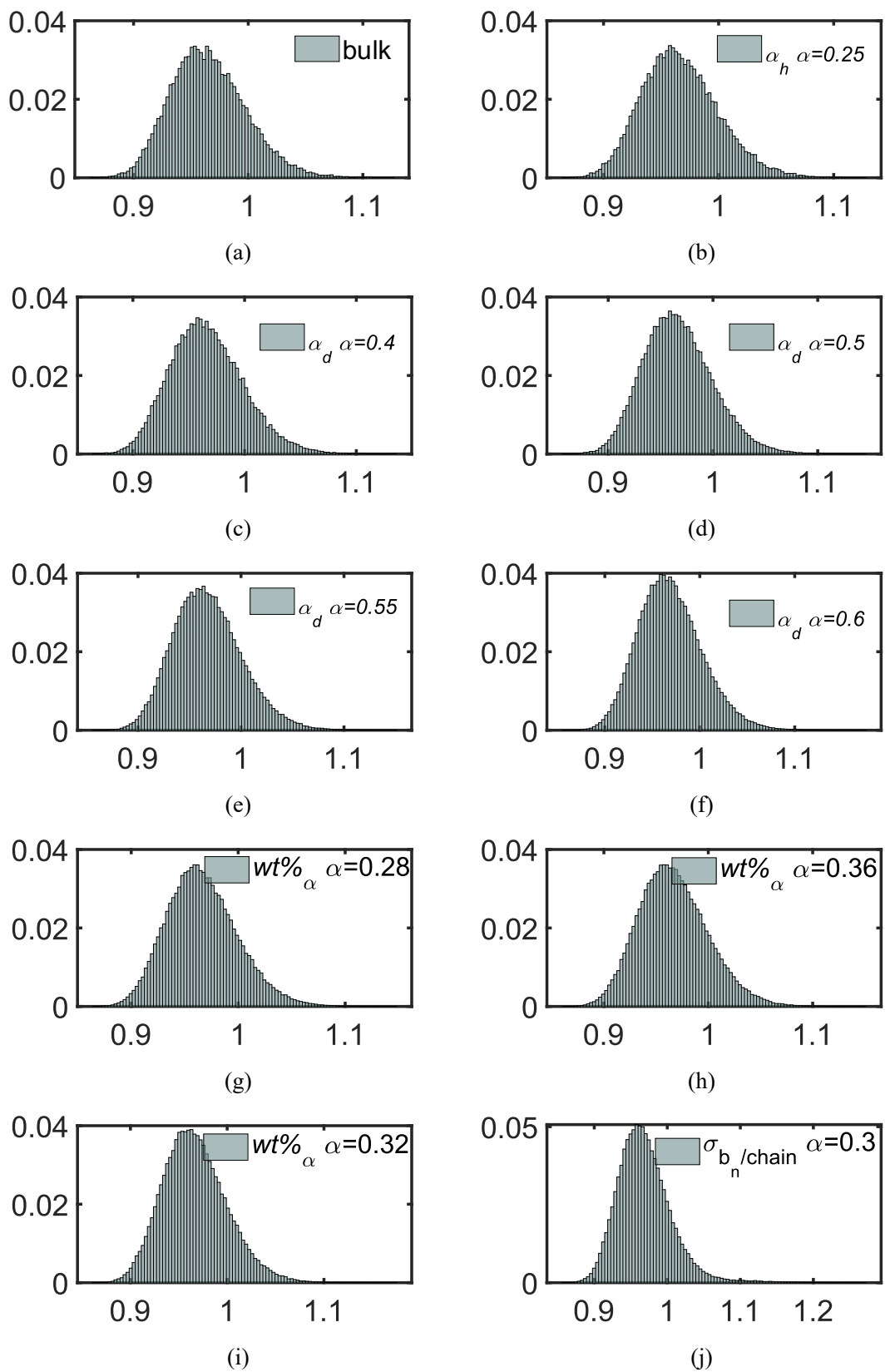

**Figure S4.** Bond distances and bond energies of polymer systems.

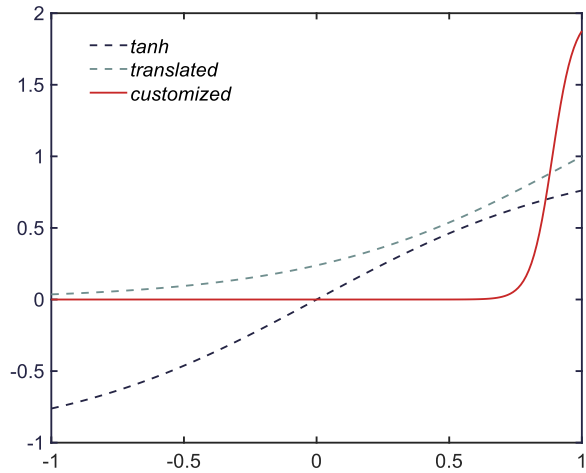

**Figure S5**. The shape of the tanh function is controlled by the two fitting parameters $p$ and $q$. Here our customized tanh function ($p = 12.2276$ and $q = 10.8695$) is plotted along with the basic tanh(x) function and translated 1+tanh(x-1) function. Customized 1+tanh(px-q) function to illustrate the conformational change induced hardening effect of the thin crosslinked polymer.

**Table S3.** Average bond distances of our simulations. It can be seen that there is no significant difference in average bond distance among different polymer films investigated by the CGMD framework.

| Groups | Average bond distance | Difference |
| --- | --- | --- |
| Bulk | 0.9655 | - |
| $\alpha_h$, $\alpha = 0.25$ | 0.9669 | 0.14% |
| $\alpha_d$, $\alpha = 0.4$ | 0.9663 | 0.08% |
| $\alpha_d$, $\alpha = 0.55$ | 0.9667 | 0.12% |
| $\alpha_d$, $\alpha = 0.60$ | 0.9675 | 0.20% |
| $\alpha_d$, $\alpha = 0.28$ | 0.9680 | 0.25% |
| $\alpha_d$, $\alpha = 0.36$ | 0.9665 | 0.10% |
| $wt\%_d$, $\alpha = 0.4$ | 0.9667 | 0.12% |
| $\sigma_{bd}$, $\alpha = 0.3$ | 0.9667 | 0.12% |

The mechanical performance of crosslinked polymers should be affected by their connecting modes. The modulus of a polymer system with all the crosslinkers connected in parallel may correspond to an upper limit estimation; the modulus of a perfectly series system may be regarded as a lower bound; and a more general polymer network is expected to have the modulus with its value in between.

To demonstrate the effect in our scaling, the simplest way is to add coefficients of connectivity to each system considered in this work, as Equation 2 shows. An almost perfect fit could be achieved as Figure S6 shows by this method.

$$E \cong \frac{1}{n}\left[\varphi \sum_{i=1}^{n} \frac{N_i^2}{(\bar{N}r_i)^2} + Ck \sum_{i=1}^{n} \theta \left[1 + tanh\left(p\frac{r_i}{N_i} - q\right)\right]\right] \quad \text{Equation S1}$$

However, it must be noted that, in a fully crosslinked polymer system, the connection modes are also highly correlated with the conformations, for example $\alpha_d$s. Thus, this method may provide overfitted results. Detailed research into the correlation between the conformations and the connection modes will be beneficial to further understand these effects.

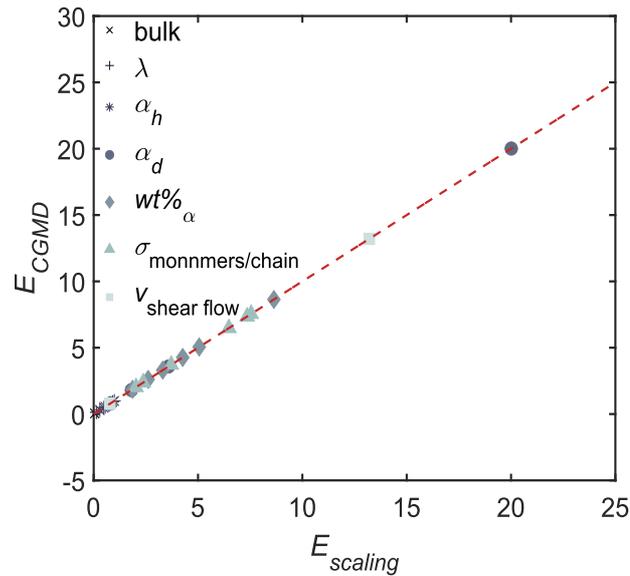

**Figure S6**. Demonstration of the series/parallel effect of the polymer chains.

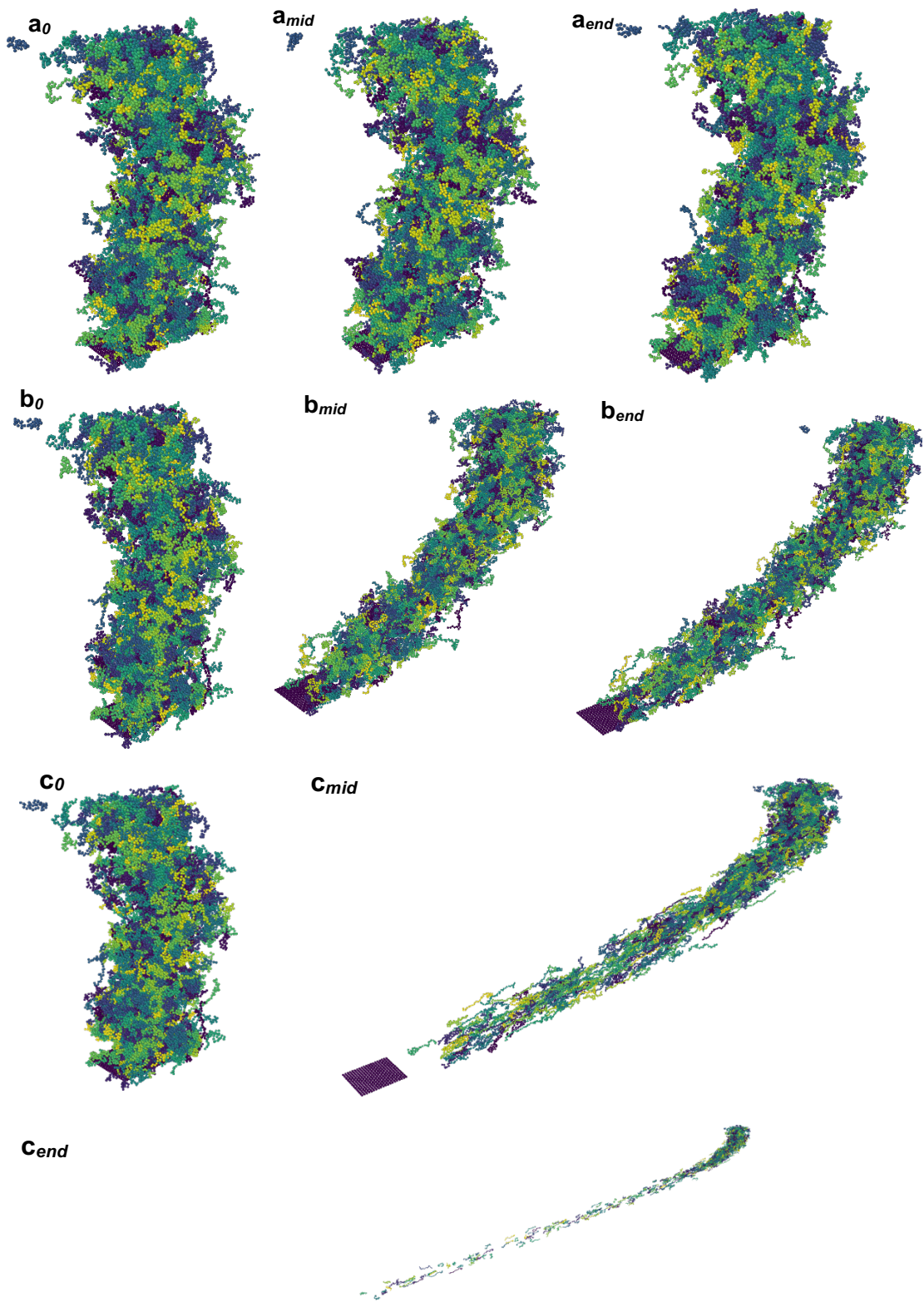

**Figure S7**. Besides the thickness of the polymer, the radial force, which positively related with the spin coating speed, also largely affect the conformation of the chains in processing. The conformational change of the polymer films under **a.** small ($1\times10^{-4}$ in unit LJ), **b.** medium ($1\times10^{-3}$ in unit LJ) and **c.** large ($1\times10^{-2}$ in unit LJ) external atom force are shown.

**Table S4.** Relations between spin coating speed, thickness of polymers, and virtual tangential speed in simulation.

| PDMS/toluene solution concentration (wt%) | spin coating speed | thickness of the polymers | virtual tangential speed |
|---|---|---|---|
| 4 | 4000 | 50 | $4\pi/3$ m/s, 1.11 in Unit LJ |
| 6 | 2000 | 150 | $2\pi/3$ m/s, 0.554 in Unit LJ |
| 10 | 2000 | 500 | |

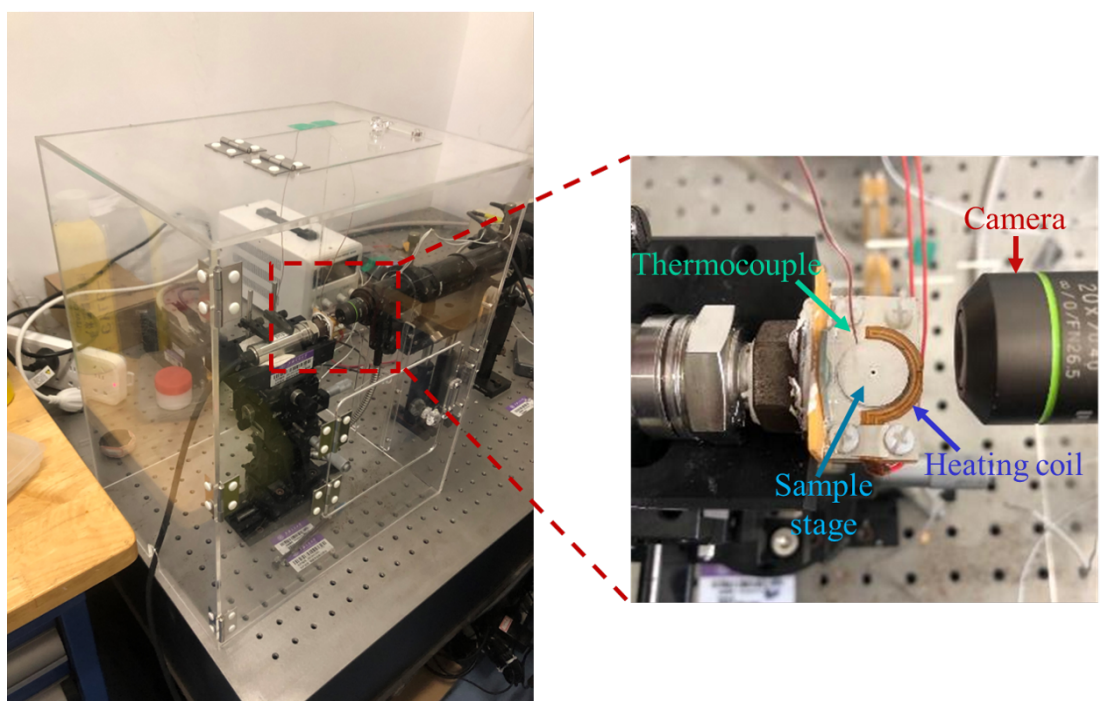

**Figure S8.** In-house micro vibration device for the mechanical response measurement of the freestanding thin polymer films.

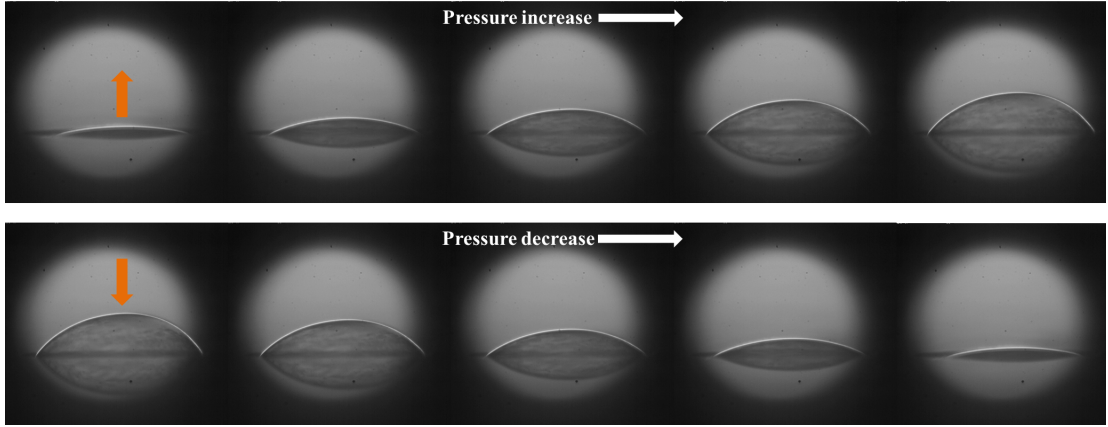

**Figure S9**. Continuous captures of the polymer film samples under one triangle pressure cycle. The first row shows the shape change of film pressure increasing, and the second row shows the process of pressure decreasing. Orange arrows indicate the deformation direction of the film.

In-plane stresses and strains are calculated from radius of curvature and pressure by the following equations:

$$\sigma_{11} = \sigma_{22} = \frac{PR}{2t} \qquad \text{Equation S2}$$

$$\epsilon_{11} = \epsilon_{22} = \frac{R\sin^{-1}(\frac{a}{R})}{a} - 1 \qquad \text{Equation S3}$$

where $P$ represents the air pressure, $R$ is the radius of curvature of PDMS polymer film obtained from photos, $t$ is the thickness of PDMS polymer film and $a$ is the radius of glass substrate hole as Figure 5g shows.